\documentclass[12pt]{article}

\newsavebox{\ns}
\newsavebox{\dbrane}
\newsavebox{\dbshort}

\usepackage{epsfig}
\usepackage{amssymb}

\def\be{\begin{eqnarray}}
\def\ee{\end{eqnarray}}

\newcommand{\nn}{\nonumber}
\newcommand\para{\paragraph{}}
\newcommand{\ft}[2]{{\textstyle\frac{#1}{#2}}}
\newcommand{\eqn}[1]{(\ref{#1})}

\def\Dslash{\,\,{\raise.15ex\hbox{/}\mkern-12mu D}}
\def\Dbarslash{\,\,{\raise.15ex\hbox{/}\mkern-12mu {\bar D}}}
\def\delslash{\,\,{\raise.15ex\hbox{/}\mkern-9mu \partial}}
\def\delbarslash{\,\,{\raise.15ex\hbox{/}\mkern-9mu {\bar\partial}}}
\def\pslash{\,\,{\raise.15ex\hbox{/}\mkern-9mu p}}
\def\calDslash{\,\,{\raise.15ex\hbox{/}\mkern-12mu {\cal D}}}

\begin{document}
\pagestyle{plain}
\setcounter{page}{1}
\newcounter{bean}
\baselineskip16pt

\begin{titlepage}

\begin{center}
\today
\hfill hep-th/0306266\\
\hfill MIT-CTP-3390 \\

\vskip 1.5 cm
{\large \bf A Quantum Hall Fluid of Vortices}
\vskip 1 cm 
{David Tong}\\
\vskip 1cm
{\sl Center for Theoretical Physics, 
Massachusetts Institute of Technology, \\ Cambridge, MA 02139, U.S.A.\\}

\end{center}

\vskip 0.5 cm
\begin{abstract}
In this note we demonstrate that vortices in a non-relativistic Chern-Simons 
theory form a quantum Hall fluid. We show that the vortex dynamics is 
controlled  by the matrix mechanics previously 
proposed by Polychronakos as a description of the quantum Hall droplet. 
As the number of vortices becomes large, they fill the plane and a 
hydrodynamic treatment becomes possible, resulting in the non-commutative 
theory of Susskind. Key to the story is the recent D-brane 
realisation of vortices and their moduli spaces.

\end{abstract}

\end{titlepage}

\subsection*{The Introduction}

Chern-Simons theories \cite{roman} provide an effective, long distance description of the 
fractional quantum Hall effect (FQHE). In fact, they provide several such 
descriptions. The range of models on the market fall roughly into one of two 
categories depending on the physical 
interpretation of the vector potential $A$ appearing in the $A\wedge F$ term. 
In the initial papers on the subject \cite{book,zhk}, $A$ acts as a statistical 
gauge field of the type suggested by Wilczek \cite{frank}. Its role is  to endow the 
excitations of the model with the charge and statistics appropriate to the 
quantum Hall system. Later works concentrate on hydrodynamic properties 
of the quantum Hall fluid in which either $A$ or ${}^\star F$ are vector fields 
associated with conserved currents and charge density 
\cite{blokwen,bs}\footnote{For readers whose brane activity usually takes place at the 
Planck scale, introductions to facets of the FQHE may be found in 
\cite{book,girvin,zee}.}.
\para
More recently, Susskind has suggested that the hydrodynamic properties of 
the quantum Hall fluid 
are captured by a Chern-Simons theory at level $k$, defined on a 
non-commutative background \cite{suss}. 
The electrons sit at Laughlin filling fraction $\nu=1/(k+1)$ and the fluid fills 
the infinite plane. 
Subsequently, Polychronakos proposed a matrix model regularisation of Susskind's 
theory in order to describe a finite quantum Hall droplet 
consisting of $N$ electrons \cite{alexios}. 
As $N\rightarrow\infty$, the droplet expands to fill the plane and we recover 
Susskind's non-commutative dynamics. Several properties of this matrix model have 
since been explored, including the relationship to Laughlin wavefunctions 
\cite{hellvram,ks} and the coupling to external electromagnetic 
fields \cite{hkk}.
\para
In this paper we study a non-relativistic Chern-Simons theory defined on an 
ordinary, mundane space in which the coordinates commute. 
The theory does not give an immediate description of a fractional quantum Hall 
fluid, but rather defines a background into which spin-polarised (i.e. spinless) 
electrons may be injected. These electrons arise as the vortices of the 
theory and we show that their quantum dynamics is controlled by the 
matrix model of Polychronakos. The vortices thus form a fractional quantum Hall droplet.
As the number of vortices becomes large, they may be described by Susskind's hydrodynamic, 
non-commutative Chern-Simons theory. 
\para
The key to the connection between vortex dynamics and the FQHE is provided by 
the recent string theory realisation of vortices and their moduli spaces given 
in \cite{me}. While \cite{me} considered vortices in the relativistic 
Maxwell-Higgs theory, we here extend the results to the non-relativistic 
Chern-Simons case. Rather 
than present a new D-brane picture, we instead make use of known connections between 
vortex dynamics in Maxwell and Chern-Simons theories \cite{manton,romao}.
\para
The observation that, under favourable conditions, vortices may form a quantum Hall fluid is 
hardly new. It is implicit in the hierarchy construction of FQH states through the 
condensation of quasiparticles. 
In the context of superconductivity, it was first suggested by Stern \cite{stern}, 
motivated by the similarity 
between the Magnus force and the Lorentz force. 
Recently the idea has received much attention 
in the context of rotating Bose systems - see for example \cite{bec}. 
Here we give a simple, string theory inspired, derivation of this effect. 

\subsection*{The Vortex}

The non-relativistic 
model that we consider consists of a single complex scalar field $\phi$, coupled 
to a $U(1)$ gauge field $a_\mu$, 
\be
L= \int d^2x\ \ i\phi^\dagger{\cal D}_0\phi-\frac{k}{4\pi}
\epsilon^{\mu\nu\rho}a_\mu \partial_\nu a_\rho-\mu a_0 
-{\Delta}|{\cal D}_i\phi|^2 
-\frac{2\pi\Delta}{k}\left(|\phi|^2-\mu\right)^2
\label{lag}\ee
where the covariant derivative is given by ${\cal D}_\mu=\partial_\mu-ia_\mu$. The 
theory was previously proposed by Manton \cite{manton} as a non-dissipative model for 
vortex motion in superconductors\footnote{Strictly speaking, the Lagrangian 
agrees with that of \cite{manton} only after imposing Gauss' law.}. Here 
$a_\mu$ will play the role of a statistical gauge field. The equation of 
motion for $a_0$ yields Gauss' law,
\be
b=\epsilon^{ij}\partial_{i}a_j=\frac{2\pi}{k}\left(|\phi|^2-\mu\right)
\label{gauss}\ee
The chemical potential term $\mu a_0$ ensures that the potential energy can 
be minimised by $|\phi|^2=\mu$ with $b=0$. The theory lives in a 
gapped phase with broken gauge symmetry, and therefore admits topologically 
stable vortices with winding number $N\in\mathbb{Z}$
\be
\int d^2x\ {b}=-2\pi N
\label{winding}\ee
The coefficients in \eqn{lag} are not arbitrary. 
Firstly, we require that the Chern-Simons level is quantised: 
$k \in{\mathbb Z}$. We pick $k>0$. Secondly, the coefficients in the 
potential energy have been fine-tuned so that the second order equations of 
motion may be integrated once \cite{jp}. 
It can be checked that, for winding number $N>0$, the equations 
of motion are satisfied by solutions to \eqn{gauss} together with the 
first order equation,
\be
{\cal D}_z\phi=0
\label{bog}\ee
where we have defined the complex coordinate $z=x^1+ix^2$ and $a_z=\ft12(a_1-ia_2)$. 
The energy required to excite $N$ such vortices is ${\cal E}=2\pi\mu N\Delta$. This 
formula contains no hint of binding energy and, indeed, it can be shown that that 
there 
are no static forces between the vortices. Note that because equation \eqn{gauss} 
comes from Gauss' law, vortices with this property exist only for $N>0$. 
\para
Although our starting point was a non-relativistic Chern-Simons theory, the 
vortex equations \eqn{gauss}, \eqn{winding} and \eqn{bog} coincide with those 
arising in the relativistic Maxwell-Higgs model. The solutions to these equations 
therefore also describe vortices in a critically coupled superconductor (i.e. on the 
borderline between type I and type II). The fact that the same vortices are shared 
by the non-relativistic 
Chern-Simons theory and the relativistic Maxwell theory will prove crucial in the 
following.
\para
While no analytic solutions to the 
vortex equations are known, index theorems reveal that the most general solution 
contains $2N$ parameters \cite{erick}. 
These may be taken to be unordered $N$-tuple of 
positions $z^a$, $a=1,\ldots,N$ on the complex plane, each of which corresponds 
to a zero of the Higgs field. The {\it moduli space} of vortices, 
defined as the space of solutions to the vortex equations, is 
therefore a $2N$-dimensional manifold which we shall denote as ${\cal M}_N$. 
Geometrically, ${\cal M}_N\cong {\bf C}^N/S_N$, where $S_N$ is the permutation group of 
$N$ elements, reflecting the fact that the vortices are indistinguishable. 
In the asymptotic region of ${\cal M}_N$, when $|z^a-z^b|$ is larger than 
all other length scales, the solution looks like $N$ well-separated vortices, each 
containing a single quantum of flux. However, as the 
vortices approach, the orbifold singularities of ${\bf C}^N/S_N$ are smoothed 
out. At this point the $z^a$ are no longer good coordinates and one 
should transform to another basis in which ${\cal M}_N$ is manifestly smooth. 
The purpose of this paper is to show that in this regime, as the vortices approach, 
they form a quantum Hall fluid.

\subsection*{The Dynamics}

The Lagrangian \eqn{lag} was chosen so that there are no static forces between 
vortices. In a derivative expansion, the velocity dependent interactions are therefore 
dominant. For slow moving vortices, 
these may be elegantly captured using the Manton moduli space approximation. This 
assumes that all time dependence is restricted to the collective coordinates $z^a=z^a(t)$. 
Substituting the time dependent configurations into the kinetic terms of \eqn{lag} then gives 
rise an effective quantum mechanics for $z^a$. 
\para
Let us first recall the story for vortices in the relativistic Maxwell-Higgs model 
\cite{samols}, since this situation will turn out to be intimately woven with our own. Here 
the kinetic terms are second order and ${\cal M}_N$ is understood as the 
{\it configuration space} of the vortex system. 
The moduli space approximation defines a K\"ahler 
metric $g$ on ${\cal M}_N$ which captures the low-energy energy dynamics,
\be
{\cal L}_{\rm Maxwell} = \ft12 g_{ab}(z^c,\bar{z}^c)\,\dot{z}^a\dot{\bar{z}}^b
\label{metric}\ee
The metric $g$ is constructed in such a way that the geodesics track the classical 
scattering of vortices. 
\para
In the present case, the kinetic terms in our non-relativistic Lagrangian 
are first order   and ${\cal M}_N$ now plays the role of the {\it phase space} 
of the vortex system. 
The low-energy dynamics of the vortices is of the form,
\be
L_{\rm CS}=  \ft{i}{2}\left(\bar{f}_a(z^b,\bar{z}^b)\, \dot{z}^a -{f}_a(z^b,\bar{z}^b)\, 
\dot{\bar{z}}^a\right)
\label{answer}\ee
where ${\cal A}=\bar{f}_adz^a-{f}_ad\bar{z}^a$ 
is a connection on ${\cal M}_N$. The task of determining 
${\cal A}$ was undertaken by Manton \cite{manton} and Rom$\tilde{\rm a}$o 
\cite{romao}. For far separated vortices, they show that 
$\bar{f}_a\rightarrow \pi\mu\bar{z}^a$ 
which simply describes non-interacting fluxes in the condensate $\mu$. In this regime, 
the Lagrangian becomes equivalent to one describing non-interacting electric charges 
in a large magnetic field $B=2\pi\mu$, providing a dual picture to which we shall 
return later. 
For the purposes of this paper we are 
more interested in the physics when the vortices approach. Here an explicit 
expression for ${\cal A}$ is not known. However, it can be shown that 
${\cal A}$ has the simple geometrical interpretation \cite{manton,romao} 
\be
d{\cal A}= -i\Omega
\label{nice}\ee
where $\Omega$ is the K\"ahler form with respect to the metric $g$ on ${\cal M}_N$. 
This result provides a connection between the dynamics of 
vortices in the Chern-Simons theory and the dynamics of vortices in the Maxwell theory, 
and will play an important role in the following section. However, it is not of immediate 
use in determining the physics of closely packed vortices. The trouble lies 
in the fact that, like ${\cal A}$, little is known about the metric $g$. 
In the asymptotic regime $|z^a-z^b|\gg 1$ the metric becomes flat, once again 
reflecting the fact that far-seperated vortices may be thought of as non-interacting particles. 
To make progress in understanding the dynamics in the limit in which the vortices 
approach, we turn to string theory for inspiration. 

\subsection*{The Matrix}

Let us start once more with vortices in 
the relativistic Maxwell-Higgs theory. Recently, a D-brane construction 
of this model was given in type IIB string theory \cite{me}. In this 
set-up, the vortices appear as D-strings suspended between NS5-branes 
and D3-branes, and their dynamics can be easily determined.
Let us quickly review the main result. 
It was found that the dynamics of 
$N$ D-strings is encoded in a $U(N)$ gauged quantum mechanics, containing a complex matrix 
$Z^a_{\ b}$, $a,b=1,\ldots,N$ transforming in the adjoint of $U(N)$, and a complex vector 
$\psi^a$ 
transforming in the fundamental representation. 
The low-energy dynamics of the D-strings is given by\footnote{We have rescaled 
$Z$ by the vortex mass relative 
to \cite{me} so that it has the correct dimension. To compare with the conventions of 
\cite{me}, note that $\zeta\equiv\mu$ and $e^2\equiv 2\pi/\kappa$.}. 
\be
{\cal L}_{D-brane}= {\rm Tr}\, \left(\pi\mu\,{\cal D}_tZ^\dagger{\cal D}_tZ
-\lambda^2\left(\psi\psi^\dagger+
\pi\mu\,[Z,Z^\dagger]-\kappa\right)^2\right)+{\cal D}_t\psi^\dagger{\cal D}_t\psi
\label{dbrane}\ee
Here ${\cal D}_tZ=\dot{Z}-i[A_0,Z]$ and ${\cal D}_t\psi=\dot{\psi}-iA_0\psi$ where  
$A_0$ is a vector potential which may be completely gauged away. 
String theory instructs us to take the $\lambda^2\rightarrow\infty$ limit, imposing 
the $N^2$ constraints  
\be
\psi^a\psi^\dagger_b+\pi\mu\,[Z,Z^\dagger]^a_{\ b}=\kappa\delta^a_{\ b}
\label{constraint}\ee
on the $2N(N+1)$ degrees of freedom contained within 
$Z$ and $\psi$. Restricting to $U(N)$ invariant 
objects as required by the gauge symmetry 
imposes a further $N^2$ constraints, leaving a remaining $2N$ degrees of 
freedom. These describe the positions of the ends of $N$ D-strings moving 
on the plane. Since the D-strings are identified with vortices, 
these $2N$ degrees of freedom given 
give natural coordinates on the moduli space ${\cal M}_N$. 
\para
The details of the classical D-brane dynamics described by the matrix model \eqn{dbrane} 
do not coincide with the vortex dynamics described by the 
 moduli space metric 
\eqn{metric}. Nevertheless, the matrix mechanics does capture many of 
the qualitative features of the vortices, including the symmetries, 
singularity structure and scale of the moduli space. Moreover, 
when attention is restricted to certain ``topological'' or 
``BPS'' quantum 
correlation functions in supersymmetric theories, one can replace the true vortex 
dynamics \eqn{metric} with the D-brane dynamics \eqn{dbrane} and obtain quantitatively 
correct answers - see \cite{me} for further discussions. 
\para
In this paper we shall describe the dynamics of vortices in the Chern-Simons theory 
in a similar matrix fashion. Without supersymmetry as our guardian, it is hard to 
rigorously justify this step. 
Nevertheless, we continue forward under the assumption that the matrix mechanics correctly 
captures the relevant qualitative features of the vortex moduli space. 
Given the conclusions of this paper, it would be interesting to return to the moduli 
space description \eqn{answer}, 
perhaps using the geometric quantisation techniques propounded in \cite{romao}, 
in an attempt to reproduce the results without resorting to string theory. 
\para
So, if the matrix model \eqn{dbrane} describes the dynamics of vortices in the 
Maxwell theory, what is the relevant matrix model to describe the dynamics 
of vortices in our Chern-Simons theory? The answer 
lies in the relationship $d{\cal A}=-i\Omega$ which relates the vortex dynamics in 
the two theories. 
We must simply ensure that our two matrix models obey a similar relationship.
To do this, we first 
need an expression for the counterpart of $\Omega$ in the matrix model. In 
fact, this is rather simple since the matrix model is constructed in such a 
way that the K\"ahler form on ${\cal M}_N$ is inherited from the canonical 
K\"ahler form on the unconstrained space  parameterised by $Z$ and $\psi$. 
This process, known as the symplectic quotient construction, ensures that 
we may work with the obvious first order system using the variables $Z$ and $\psi$,
\be
{\cal L}= i\pi\mu\,{\rm Tr}\left(Z^\dagger\dot{Z}\right)+i\psi^\dagger\dot{\psi}
\nn\ee
 and subsequently restrict to the moduli space ${\cal M}_N$ defined by $U(N)$ 
invariant observables subject to the constraint \eqn{constraint}. This latter step may  
be achieved by 
re-introducing $A_0$, now playing the role of a Lagrange multiplier. The 
low-energy dynamics of the vortices may therefore be described by the matrix mechanics
\be
{\cal L}_{\rm matrix}= {\rm Tr}\left(i\pi\mu\,Z^\dagger{\cal D}_t{Z} - \kappa A_0\right) 
+i \psi^\dagger{\cal D}_t\psi
\label{ohbaby}\ee
This expression, describing the dynamics of Chern-Simons vortices, 
is the main result of this paper.

\subsection*{The Hall Fluid}

The matrix model \eqn{ohbaby} was previously proposed by Polychronakos as  
a description of $N$ electrons moving in the lowest Landau level of a background 
magnetic field $B=2\pi\mu$ \cite{alexios}. 
The electrons are identified with our vortices, and from now on we treat the 
terms synonymously. The classical and quantum dynamics arising from the matrix model 
have been studied in great detail (see \cite{alexios} and references therein). 
Here we mention a few choice details.  Most pertinently, it can be shown that 
when the electrons coalesce, they manifest the properties of a quantum Hall fluid of 
density $\rho$ where
\be
B=2\pi\mu\ \ \ \ ,\ \ \ \ \rho=\frac{\mu}{k}
\nn\ee
This gives rise to a classical filling fraction $\nu=2\pi\rho/B=1/k$. 
In fact, there is an important quantum shift  pointed out in \cite{alexios} 
(see also \cite{hellsuss}) so that the system actually describes a Hall fluid at 
filling fraction $\nu=1/(k+1)$. 
\para
Let us try to understand this behaviour from the perspective of critically coupled 
vortices. At first glance 
it seems peculiar that the vortex dynamics would give rise to a 
FQH fluid since the Lagrangian \eqn{answer} contains no sign of the repulsive particle 
interactions that are usually held accountable for such an effect. Indeed, the 
Hamiltonian associated to \eqn{answer} vanishes and, for $|z^a-z^b|$ suitably large, 
the solution to the vortex equations can be understood as $N$ far-separated, 
non-interacting vortices. Each has size $L\sim\sqrt{k/\mu}$ 
which (ignoring factors of $2$ and $\pi$) is the penetration depth in the 
language of superconductivity. In this paper we are interested in the 
the situation with $|z^a-z^b|<L$. What do the vortex configurations  
look like in this regime? We suggest that the 
vortices should not be thought of as overlapping particles,  
but rather as a {\it classically} incompressible fluid whose density remains 
constant at $L^{-2}$ for all values of $|z^a-z^b|<L$. 
To see that this gives rise to a consistent picture, note that the vortices see 
a background condensate $\mu$ 
which, as we have seen, can be thought of as a background magnetic flux for 
charged particles in a dual picture. The density of vortex states required to fill the 
"dual Landau level'' is therefore $\sim\mu$. With the vortices at a density of 
$L^{-2}$, this gives rise to the required filling fraction $\nu\sim 1/\mu L^2\sim 1/k$.
Clearly the speculations offered in this paragraph refer to properties of the 
classical vortex solutions, and it is to be hoped that they can be confirmed (or 
dismissed) by an explicit study of the vortex equations. 
\para
Finally, recall that as the number of electrons/vortices becomes large and 
$N\rightarrow\infty$, the constant term in the constraint \eqn{constraint} may be 
absorbed by the commutator rather than the $\psi\psi^\dagger$ term, 
\be
[Z,Z^\dagger] = \frac{k}{\pi\mu}\equiv 2\theta
\label{noncom}\ee
Expanding around this background, the matrix model \eqn{ohbaby} may be 
re-written as a $U(1)$ Chern-Simons theory at level $k$ defined on the 
plane with the coordinates satisfying \eqn{noncom}. This is Susskind's 
hydrodynamic description of the FQHE \cite{suss}. It is amusing that, having 
started  with a commutative $U(1)$ Chern-Simons-Higgs theory at level $k$, we return 
via vortex dynamics to a non-commutative $U(1)$ Chern-Simons theory at level $k$. 

\subsection*{The Potential}

As it stands, there is nothing to keep the electrons in \eqn{ohbaby} from wandering 
over the plane. When the electrons coalesce they form a FQH fluid, but when 
they sit far apart they return to their individual, yet indistinguishable, 
electronic identities. In order to energetically distinguish these two scenarios 
and coax the electrons together, 
Polychronakos introduced a simple harmonic oscillator potential \cite{alexios} 
whose role is 
to trap the electrons close to the origin,
\be
V=\frac{Bw}{2}{\rm Tr}\left(Z^\dagger Z\right)
\label{pot}\ee
In this section, we will see how to generate such a potential for the vortex 
dynamics of the Chern-Simons theory. First note that if our only requirement is 
to provide a rotationally symmetric potential which will be seen by the vortices 
and pen them near the origin, 
then one could simply add to \eqn{lag} a term of the form
\be
V_0= \frac{Bw}{2}\int d^2x\ |z|^2\left(\frac{k}{2\pi}b^2+|{\cal D}_i\phi|^2\right)
\nn\ee
However, if we want to match to the energetics of \cite{alexios}, then it 
is possible to provide the deformation that gives rise 
to the harmonic oscillator potential \eqn{pot}. The key observation is that 
equation \eqn{pot} is a mass term for $Z$ which 
induces a potential on the moduli space ${\cal M}_N$ 
that is (up to an unimportant constant) proportional to the norm-squared of the 
Killing vector associated to rotational symmetries. Such potentials appear 
frequently in soliton dynamics and can be 
written as the overlap of the corresponding zero modes of the soliton using the 
method of \cite{meagain}. 
Here we omit the details (mostly associated with gauge fixing the zero mode) and 
simply state the result: the potential
\eqn{pot} for the vortex dynamics is generated by augmenting the Chern-Simons 
Lagrangian with the potential
\be
V=V_0+ \frac{Bwk}{4\pi} \int d^2x\ \frac{1}{2}(\partial_i\Lambda)^2 + 
\frac{2\pi}{k}\Lambda^2|\phi|^2 -2\Lambda b
\nn\ee
Here the function 
$\Lambda$ arises when fixing the gauge for the vortex zero mode 
and is to be evaluated on the solution to its classical equation 
of motion in the background of the vortex.

\subsection*{The End}

Let us mention a few generalisations of the story. 
The Lagrangian of our Chern-Simons theory \eqn{lag} was fine-tuned to ensure that the vortices 
experience no static force. It is natural to wonder what happens if this is no 
longer the case. For example, we may change the coefficient of the potential term 
in \eqn{lag} by 
adding,
\be
\Delta V= \gamma\int d^2 x \left(|\phi|^2-\mu\right)^2
\nn\ee
Then for $\gamma<0$, the vortices attract (type I superconductivity), 
while for $\gamma>0$, the vortices repel (type II). In this latter 
case, the repulsive force competes with the harmonic oscillator 
potential \eqn{pot} which pushes the vortices towards the origin. For small 
$\gamma$, we expect the quantum Hall state to persist. In contrast, for suitably 
large 
$\gamma$ the vortices will undergo a phase transition to the 
more familiar Abrikosov lattice or, in the dual language of electrons, the Wigner 
crystal.
\para
The D-brane construction of \cite{me} provides several further generalisations, 
including non-Abelian Chern-Simons terms, extra scalar fields, and non-commutative 
backgrounds. For example, one could consider 
vortices in $U(m)$ Chern-Simons-Higgs theory. The 
low-energy dynamics of these vortices is described by the matrix model \eqn{ohbaby}, 
now  with $m$ vectors $\psi$. This model describes $m$ quantum Hall layers and was 
previously studied in \cite{morpoly}. As the number of vortices becomes large, 
it reduces to $U(m)$ non-commutative Chern-Simons theory. 
\para
To summarise, we have shown that the fractional quantum Hall matrix model 
of Polychronakos \cite{alexios} 
can be thought of as describing the low-energy dynamics of vortices 
in a non-relativistic Chern-Simons theory. We suggest that the physical reason 
for this behaviour is the classically incompressible nature of vortices as they 
coalesce. A crucial ingredient in our story was the D-brane construction 
of \cite{me} and, in the absence of a field theory derivation, the quantum Hall 
fluid of critically coupled vortices can be taken as a prediction of string theory. 
It is to be hoped that this new perspective on the quantum Hall matrix model may help 
in building the dictionary 
to physical quantities such as currents, particle density and the Laughlin wavefunctions.

\subsection*{The Acknowledgments}
I'm extremely grateful to Alexios Polychronakos, Jan Troost and Ashvin 
Vishwanath, each of whom spent many hours patiently listening and explaining. 
I'd also like to thank Ami Hanany and Arun Paramekanti for useful discussions. 
I'm supported by a Pappalardo fellowship, and would like to thank  
the Pappalardo family for their generosity. This work was also supported in part 
by funds provided by the U.S. Department of Energy (D.O.E.) under 
cooperative research agreement \#DF-FC02-94ER40818.


\begin{thebibliography}{99}

\small
\parskip=0pt plus 2pt

\bibitem{roman} S.~Deser, R.~Jackiw and S.~Templeton,
``{\em Topologically Massive Gauge Theories}''
Annals Phys.\  {\bf 140}, 372 (1982)

\bibitem{book}
``{\em The Quantum Hall Effect}'', edited by R. E. Prange and 
S.M Girvin. (Springer-Verlg, New York, 1986). See S. M. Girvin, Chap 10.

\bibitem{zhk}
S.~C.~Zhang, T.~H.~Hansson and S.~Kivelson,
``{\em An Effective Field Theory Model For The Fractional Quantum Hall Effect}''
Phys.\ Rev.\ Lett.\  {\bf 62} (1988) 82.

\bibitem{frank}
F.~Wilczek,
``{\em Quantum Mechanics Of Fractional Spin Particles}''
Phys.\ Rev.\ Lett.\  {\bf 49}, 957 (1982).

\bibitem{blokwen}
B.~Blok and X.~G.~Wen,
``{\em Effective Theories Of Fractional Quantum Hall Effect At Generic Filling Fractions}''
Phys.\ Rev.\ B {\bf 42} (1990) 8133.


\bibitem{bs}
S.~Bahcall and L.~Susskind,
``{\em Fluid Dynamics, Chern-Simons Theory And The Quantum Hall Effect}'',
Int.\ J.\ Mod.\ Phys.\ B {\bf 5}, 2735 (1991).

\bibitem{girvin} 
S. M. Girvin, ``{\em The Quantum Hall Effect: Novel Excitations and Broken Symmetries}'', 
arXiv:cond-mat/9907022.

\bibitem{zee}
A. Zee, ``{\em Quantum Hall Fluids}'', arXiv:cond-mat/9501022

\bibitem{suss}
L.~Susskind,
``{\em The quantum Hall fluid and non-commutative Chern Simons theory}''
arXiv:hep-th/0101029.

\bibitem{alexios}
A.~P.~Polychronakos,
``{\em Quantum Hall states as matrix Chern-Simons theory}''
JHEP {\bf 0104}, 011 (2001)
[arXiv:hep-th/0103013].


\bibitem{hellvram}
S.~Hellerman and M.~Van Raamsdonk,
``{\em Quantum Hall physics equals noncommutative field theory}''
JHEP {\bf 0110}, 039 (2001)
[arXiv:hep-th/0103179].

\bibitem{ks}
D.~Karabali and B.~Sakita, 
``{\em Chern-Simons matrix model: Coherent states and relation to Laughlin  wavefunctions}'', 
Phys.\ Rev.\ B {\bf 64}, 245316 (2001)
[arXiv:hep-th/0106016]. \\ 
D.~Karabali and B.~Sakita,
``{\em Orthogonal basis for the energy eigenfunctions of the Chern-Simons  matrix model}''
Phys.\ Rev.\ B {\bf 65}, 075304 (2002)
[arXiv:hep-th/0107168].

\bibitem{hkk} T. H. Hansson, J. Kailasvuori and A. Karlhede, 
``{\em Charge and current in the quantum Hall matrix model}'', arXiv:cond-mat/0304271.

\bibitem{me} A.~Hanany and D.~Tong,
``{\em Vortices, Instantons and Branes}''
arXiv:hep-th/0306150.

\bibitem{manton}
N.~S.~Manton,
``{\em First order vortex dynamics}''
Annals Phys.\  {\bf 256}, 114 (1997)
[arXiv:hep-th/9701027].

\bibitem{romao}
N.~M.~Romao,
``{\em Quantum Chern-Simons vortices on a sphere}''
J.\ Math.\ Phys.\  {\bf 42}, 3445 (2001)
[arXiv:hep-th/0010277].

\bibitem{stern} A. Stern,  ``{\em Quantum Hall fluid of vortices in a 
two-dimensional array of Josephson junctions}'', 
Phys. Rev. {\bf B50}, 10092  (1994), [arXiv:cond-mat/9403017].

\bibitem{bec} N.K. Wilkin, J.M.F. Gunn, ``{\em 
Condensation of `composite bosons' in a rotating BEC}'',  Phys. Rev. Lett 
{\bf 84} 6 (2000), [arXiv:cond-mat/9906282]. \\
U. Fischer, P. Fedichev and A. Recati, 
``{\em Vortex liquids and vortex quantum Hall states in trapped rotating Bose gases}'', 
arXiv:cond-mat/0212419



\bibitem{jp}
R.~Jackiw and S.~Y.~Pi,
``{\em Soliton Solutions To The Gauged Nonlinear Schrodinger Equation On The Plane}''
Phys.\ Rev.\ Lett.\  {\bf 64}, 2969 (1990).

\bibitem{erick}
E.~J.~Weinberg,
Phys.\ Rev.\ D {\bf 19}, 3008 (1979).


\bibitem{samols}
T.~M.~Samols,
``{\em Vortex Scattering}''
Commun.\ Math.\ Phys.\  {\bf 145}, 149 (1992).

\bibitem{hellsuss}
S.~Hellerman and L.~Susskind,
``{\em Realizing the quantum Hall system in string theory}''
arXiv:hep-th/0107200.

\bibitem{meagain}
D. Tong, "{\em A note on 1/4 BPS states}", 
Phys.\ Lett.\ B {\bf 460}, 295 (1999)
[arXiv:hep-th/9902005].


\bibitem{morpoly}
B.~Morariu and A.~P.~Polychronakos,
``{\em Finite noncommutative Chern-Simons with a Wilson line and the quantum  Hall effect}''
JHEP {\bf 0107}, 006 (2001)
[arXiv:hep-th/0106072].




\end{thebibliography}
\end{document}